\documentclass[
    ,final            
  ]
  {aipproc}

\layoutstyle{6x9}

\begin{document}

\title{The Gamma Ray Burst section of the White Paper on the Status and Future of Very High Energy Gamma Ray Astronomy: \\{\it{A Brief Preliminary Report}}}


\author{A.~D.~Falcone}{
address={Pennsylvania State University, 525 Davey Lab, University Park, PA 16802}
}

\author{D.~A.~Williams}{
address={Santa Cruz Institute for Particle Physics, University of California, Santa Cruz, CA 95064}
}
\author{M.~G.~Baring}{
address={Department of Physics and Astronomy, Rice University, Houston, Texas 77251-1892}
}
\author{R.~Blandford}{
address={KIPAC, Stanford Linear Accelerator Center, Stanford University, Stanford, CA 94309}
}
\author{V.~Connaughton}{
address={National Space Science and Technology Center, Huntsville, AL 35805}
}
\author{P.~Coppi}{
address={Department of Astronomy, Yale University, New Haven, CT 06520-8101}
}
\author{C.~Dermer}{
address={US Naval Research Laboratory, 4555 Overlook SW, Washington, DC, 20375-5352}
}
\author{B.~Dingus}{
address={Los Alamos National Lab, Los Alamos, NM 87545}
}
\author{C.~Fryer}{
address={Los Alamos National Lab, Los Alamos, NM 87545}
}
\author{N.~Gehrels}{
  address={NASA/Goddard Space Flight Center, Greenbelt, MD 20771}
}
\author{J.~Granot}{
address={KIPAC, Stanford Linear Accelerator Center, Stanford University, Stanford, CA 94309}
}
\author{D.~Horan}{
address={Argonne National Lab, Argonne IL 60439}
}
\author{J.~I.~Katz}{
address={Washington University, St. Louis, MO 63130}
}
\author{K.~Kuehn}{
address={Department of Physics, Ohio State University, Columbus, OH 43210}
}
\author{P.~M\'{e}sz\'{a}ros}{
address={Pennsylvania State University, 525 Davey Lab, University Park, PA 16802}
}
\author{J.~Norris}{
address={NASA/Ames Research Center, Moffett Field, CA 94035}
}
\author{P.~Saz~Parkinson}{
address={Santa Cruz Institute for Particle Physics, University of California, Santa Cruz, CA 95064}
}
\author{A.~Pe'er}{
address={Space Science Telescope Institute, Baltimore, MD 21218}
}
\author{E.~Ramirez-Ruiz}{
address={Department of Astronomy, University of California, Santa Cruz, CA 95064}
}
\author{S.~Razzaque}{
address={US Naval Research Laboratory, 4555 Overlook SW, Washington, DC, 20375-5352}
}
\author{X.~Wang}{
address={Pennsylvania State University, 525 Davey Lab, University Park, PA 16802}
}
\author{B.~Zhang}{
address={Department of Physics and Astronomy, University of Nevada, Las Vegas, NV 89154-4002}
}

\begin{abstract}
This is a short report on the preliminary findings of the gamma ray burst (GRB) working group for the white paper on the status and future of very high energy (VHE; >50 GeV) gamma-ray astronomy. The white paper discusses the status of past and current attempts to observe GRBs at GeV-TeV energies, including a handful of low-significance, {\it{possible}} detections. The white paper concentrates on the potential of future ground-based gamma-ray experiments to observe the highest energy emission ever recorded for GRBs, particularly for those that are nearby and have high Lorentz factors in the GRB jet. It is clear that the detection of VHE emission would have strong implications for GRB models, as well as cosmic ray origin. In particular, the extended emission phase (including both afterglow emission and possible flaring) of nearby long GRBs could provide the best possibility for detection. The difficult-to-obtain observations during the prompt phase of nearby long GRBs and short GRBs could also provide particularly strong constraints on the opacity and bulk Lorentz factors surrounding the acceleration site. The synergy with upcoming and existing observatories will, of course, be critical for both identification of GRBs and for multiwavelength/multimessenger studies.
 
\end{abstract}

\maketitle


\section{Introduction}
While VHE gamma-ray astronomy is a relatively young and undeveloped field, it is currently going through a very exciting epoch, with high sensitivity telescopes detecting many new sources. In an effort to evaluate this rapid progress and to prepare for the next generation, the Division of Astrophysics of the American Physical Society requested a white paper about the status and future of ground based gamma-ray astronomy. Five science working groups and one technology working group were formed. These working groups were charged with the tasks of defining the current status of their respective fields and the scientific goals that may be addressed by future instruments.  

This report is a very brief summary of the preliminary findings of the GRB science working group for this white paper. For more detail, the reader should refer to the forthcoming comprehensive white paper.

\section{Brief Theory Overview}
Gamma-ray burst ${\nu}F_{\nu}$ spectra have a peak at photon energies ranging from a few keV to several MeV, and the spectra are nonthermal. There are several models for conversion of explosion energy including blast waves with internal+external shocks, Poynting flux dominated models, and external shocks in a clumpy ambient environment. The most commonly invoked mechanism is a relativistic jet that leads to internal shock collisions, with associated prompt emission, followed by a forward shock expanding into the ambient medium causing an extended afterglow. Both of these mechanisms lead to acceleration of particles, with associated emission, including e$^-$ synchrotron at keV to MeV energies and inverse Compton at higher energies. Protons should also be accelerated (maybe to UHE), with associated p$^+$ synchrotron and p$^+$ induced cascades. Knowledge of the high energy spectra is critical to understanding the relative importance of these processes and to understanding the GRB within its surroundings. For a more comprehensive GRB theory review, see \cite{meszaros06, PanaitescuMeszaros98, dermer07, ReesMeszaros94, PeerWaxman04b, Paczynski86}, and references therein.

Of particular importance to VHE gamma ray studies is the fact that pair production interactions of gamma rays with the IR photons of the extragalactic background attenuate the gamma-ray signal \cite{Primacketal2005}. This limits the distance over which VHE gamma rays can propagate to z $< \sim$0.5 for the current generation of telescopes.

\section{Current Status of VHE Observations}
From EGRET data, it is clear that GRB spectra extend to at least tens of GeV. Although many authors have predicted the existence of VHE emission from GRBs either during the prompt phase or at any time during the multi-component afterglow, no definitive detections have been made. Telescopes that can observe gamma ray emission above 100 GeV fall into two broad categories, air shower arrays (including water Cherenkov detectors such as Milagro) and imaging atmospheric Cherenkov telescopes (IACTs), such as HESS, VERITAS, and MAGIC. There is a possible detection in the TeV range by Milagrito \cite{Atkinsetal00,Atkinsetal03}. Stacked analyses of many satellite triggered bursts have also yielded only statistically marginal positive excess \cite{Amenomori01}. IACTs have slewed to a handful of GRBs quickly (response time ranges from ~40 s to hours), resulting in upper limits \cite{Horanetal2008, Albertetal2006}. Unfortunately, at this time, most observations with sensitive IACTs have not been prompt slews, but late-time flux limits as low as 2\% of the Crab have been reported.

\section{Potential VHE Emission}
\begin{figure}
\centering
\includegraphics[scale=0.75]{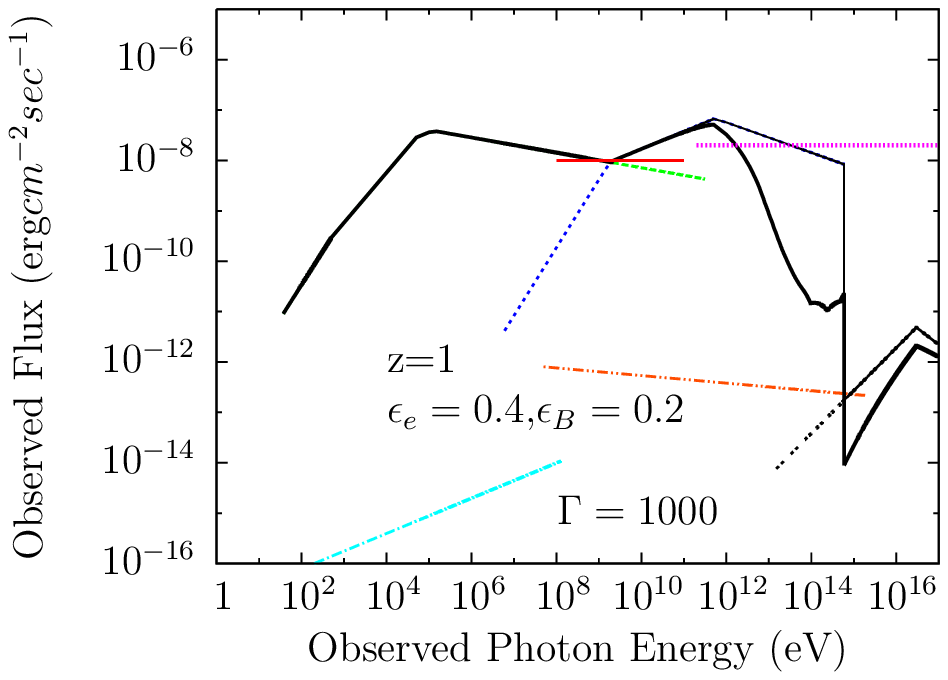}\quad
\includegraphics[scale=0.75]{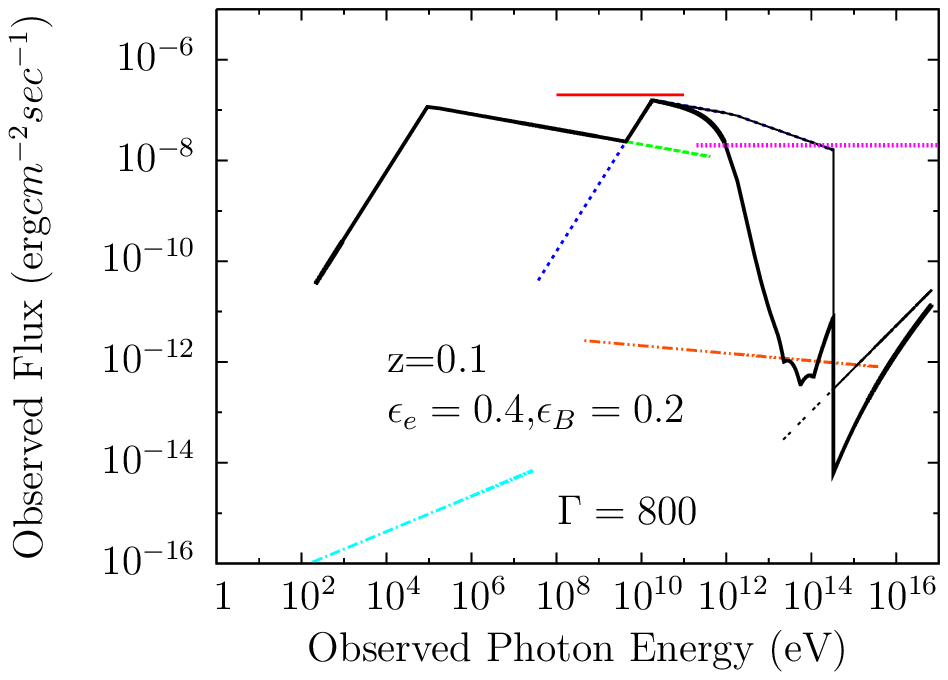}
\caption{Modelled broad-band spectrum of the GRB internal shock prompt emission \cite{GuptaZhang07}. (a) A long GRB with the observed sub-MeV luminosity of $\sim 10^{51}~{\rm erg~s^{-1}}$ is modelled. The solid black lines represent the final spectrum before (thin line) and after (thick line) inclusion of internal absorption. The long dashed green line is the electron synchrotron component; the short-dashed blue line is the electron IC component; the double short-dashed black curve is the $\pi^0$ decay component; the triple short-dashed line represents the synchrotron radiation produced by e$^\pm$ from $\pi^\pm$ decays; the dash-dotted (light blue) line represents proton synchrotron. (b) The analogous spectrum of a bright short GRB with $E_{iso}=10^{51}$erg.} 
\label{prompt}
\end{figure}

\begin{figure}
\centering
\includegraphics[width=0.40\linewidth]{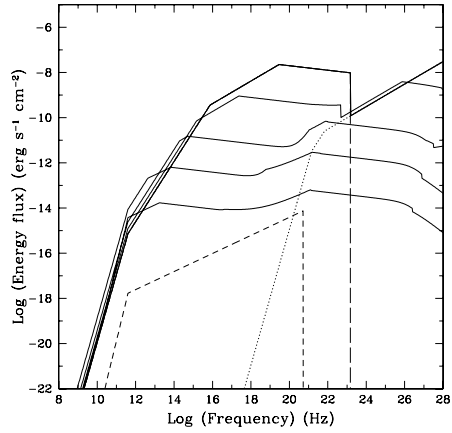}
\caption{The SSC emission from the forward shock region in the afterglow phase. Temporal evolution of the theoretical models for synchrotron and SSC components for $\epsilon_e=0.5$, $\epsilon_B=0.01$; solid curves from top to bottom are at onset, 1 min, 1 hour, 1 day, 1 month. The contributions to the emission at onset are shown as long-dashed (electron-synchrotron), short-dashed (proton-synchrotron) and dotted (electron IC) curves \cite{ZhangMeszaros01}.}
\label{afterglow}
\end{figure}

A model prediction for long GRB prompt emission, including both inverse Compton and hadronic emission, is shown in Figure~\ref{prompt}a \cite{GuptaZhang07}. Internal opacity from pair production will attenuate VHE photons, thus providing a probe of emission radius and bulk Lorentz factor while also limiting detection probability. During the afterglow phase, inverse Compton emission from forward shock e$^-$ scattering from myriad photon fields, including self synchrotron and reverse shock emission, will lead to a high energy component that may be detectable by VHE instruments. Hadronic emission should also be present. A model time-evolved spectrum is shown in Figure~\ref{afterglow}.

Short GRBs are generally dimmer than long GRBs, but they are closer, which means that a larger fraction of short GRBs will have minimal attenuation from extragalactic infrared photons. A sample prediction is shown in Figure~\ref{prompt}b.

\begin{figure}
\centering
\includegraphics[width=0.35\linewidth,angle=270]{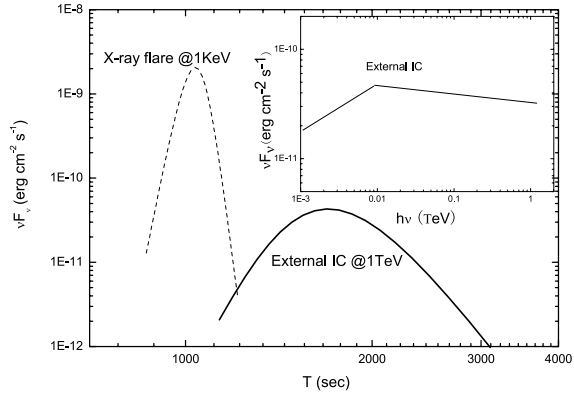}
\caption{A sample model prediction of VHE gamma rays from flares, using only an inverse Compton component \cite{wan06}.
\label{flare}}
\end{figure}

Recently, Swift has provided the remarkable result that late time (as late as $10^5$ sec) X-ray flares are a common occurrence, detected in $\sim$50\% of GRBs \cite{bur05,fal07,chi07}. It is possible that these flares have a higher energy component, potentially due to inverse-Compton scattering as predicted for a fairly modest flare in Figure~\ref{flare}. The average flare fluence is a factor of $\sim$10$\times$ less than the average prompt fluence \cite{fal07}, but there are some notable examples in which flare fluence is approximately the same as prompt fluence \cite{fal06}.

\section{Synergy with Existing and Future Instruments}
While GRB triggers are possible from wide angle VHE instruments, a space-based GRB detector will be needed. Swift, GLAST, or future WFOV hard X-ray monitors must provide low energy observations. GRBs with observations by both GLAST and VHE telescopes will be particularly exciting and may probe high Lorentz factors. Neutrino telescopes, UHECR telescopes (eg AUGER), and next generation VHE observatories can supplement one another in the search for UHECRs from GRBs since neutrinos are expected along with GeV-TeV gamma rays. A TeV trigger could also effectively improve neutrino sensitivity.

\section{Conclusions}
The challenge of VHE astronomy for contributing to GRB prompt-phase science is twofold: (i) the GRB source must be at relatively small redshifts (z < $\sim$0.5) to avoid significant attenuation of TeV radiation by the extragalactic background light and the GRB parameters must be favorable to avoid excessive internal absorption; and (ii) to insure a reasonable chance of detection, the instrument must be very sensitive with all-sky coverage or it must be very sensitive with rapid slewing ( 5-10deg/sec).

The detection of VHE afterglow emission, delayed prompt emission, and/or flare emission simply requires a sensitive instrument; with only moderate slew speed. It is likely that any instrument with $\sim$10$\times$ sensitivity improvements over the current generation of IACTs will detect GRB-related VHE emission. These instrument requirements could be achieved by next generation instruments, given sufficient resources.

Either of these detection channels would make great strides towards understanding the extreme nature and environments of GRBs, particularly the local opacity and the bulk Lorentz factor. It could also contribute to our understanding of the decades old ultra high energy cosmic ray acceleration problem, as well as GRB progenitor studies, star formation studies, and Lorentz invariance violation studies. For a more detailed report please refer to the forthcoming white paper when it is released.


\begin{thebibliography}{9}

\bibitem[Albert et al. (2006)]{Albertetal2006}
Albert, J., et al., \emph{Astrophys. Journ. Lett.}, \textbf{641}, L9

\bibitem[Amenomori et al. (2001)]{Amenomori01}
Amenomori, M., et al., AIP Conf. Proc., \textbf{558}, 844 (2001).

\bibitem[Atkins et al. (2000)]{Atkinsetal00}
Atkins, R., et al., \emph{Astrophys. Journ. Lett.}, \textbf{533}, L119 (2000).

\bibitem[Atkins et al. (2003)]{Atkinsetal03}
Atkins, R., et al., \emph{Astrophys. Journ.}, \textbf{583}, 824 (2003).

\bibitem[Burrows et~al.\ (2005)]{bur05} 
Burrows, D.N., Romano, P., Falcone, A., et~al., \emph{Science}, \textbf{309}, Issue 5742, 1833 (2005).

\bibitem[Chincarini et~al.\ (2007)]{chi07} 
Chincarini, G., et~al., \emph{Astrophys. Journ.}, \textbf{671}, 1903 (2007).

\bibitem[Dermer (2007)]{dermer07}
Dermer, C.~D., \emph{Astrophys. Journ.}, \textbf{664}, 384 (2007).

\bibitem[Falcone et~al.\ (2006)]{fal06} 
Falcone, A., et~al., \emph{Astrophys. Journ.},\textbf{641}, 1010 (2006).

\bibitem[Falcone et~al.\ (2007)]{fal07} 
Falcone, A., et~al., \emph{Astrophys. Journ.}, \textbf{671}, 1921 (2007).

\bibitem[Gupta \& Zhang (2007)]{GuptaZhang07} 
Gupta, N., Zhang, B., \emph{Mon.\ Not.\ Roy.\ Astron.\ Soc.}, \textbf{380}, 78 (2007).

\bibitem[Horan et al. (2008)]{Horanetal2008}
Horan, D., et al., \emph{these proceedings} (2008).

\bibitem[Meszaros (2006)]{meszaros06}
M\'esz\'aros, P., \emph{Rev.\ Prog.\ Phys.}, \textbf{69}, 2259 (2006).

\bibitem[Paczy\'nski (1986)]{Paczynski86}
Paczy\'nski, B., \emph{Astrophys.\ Journ.}, \textbf{308}, L43 (1986).

\bibitem[Panaitescu \& M\'esz\'aros (1998)]{PanaitescuMeszaros98}
Panaitescu, A. \& M\'esz\'aros, P., \emph{Astrophys. Journ.}, \textbf{492}, 683 (1998).

\bibitem[Pe'er \& Waxman (2004)]{PeerWaxman04b}
Pe'er, A., \& Waxman, E., \emph{Astrophys. Journ.}, \textbf{613}, 448 (2004).

\bibitem[Primack et al. (2005)]{Primacketal2005}
Primack, J.~R.; Bullock, J.~S.; Somerville, R.~S., AIP Conf. Proc., \textbf{745}, 23 (2005); astro-ph/0502177

\bibitem[Rees \& M\'esz\'aros (1994)]{ReesMeszaros94}
Rees, M.~J., \& M\'esz\'aros, P., \emph{Astrophys. Journ.}, \textbf{430}, L93 (1994).

\bibitem[Wang et~al.\ (2006)]{wan06}
Wang, X.-Y., Li, Z., M\'esz\'aros, P., \emph{Astrophys. Journ. Lett.}., \textbf{641}, L89 (2006).

\bibitem[Zhang \& M\'esz\'aros (2001)]{ZhangMeszaros01} 
Zhang, B., M\'esz\'aros P., \emph{Astrophys. Journ.}, \textbf{559}, 110 (2001).

\end{thebibliography}
\end{document}